# Advanced Analysis of Temporal Data Using Fisher-Shannon Information: Theoretical Development and Application in Geosciences

Fabian Guignard[1]*, Mohamed Laib[2], Federico Amato[1] and Mikhail Kanevski[1]

[1] Faculty of Geosciences and Environment, Institute of Earth Surface Dynamics, University of Lausanne, Lausanne, Switzerland, [2] Department of Information Technologies for Innovative Services, Luxembourg Institute of Science and Technology—LIST, Belvaux, Luxembourg

Complex non-linear time series are ubiquitous in geosciences. Quantifying complexity and non-stationarity of these data is a challenging task, and advanced complexity-based exploratory tool are required for understanding and visualizing such data. This paper discusses the Fisher-Shannon method, from which one can obtain a complexity measure and detect non-stationarity, as an efficient data exploration tool. The state-of-the-art studies related to the Fisher-Shannon measures are collected, and new analytical formulas for positive unimodal skewed distributions are proposed. Case studies on both synthetic and real data illustrate the usefulness of the Fisher-Shannon method, which can find application in different domains including time series discrimination and generation of times series features for clustering, modeling and forecasting. The paper is accompanied with Python and R libraries for the non-parametric estimation of the proposed measures.

**Keywords:** Fisher-Shannon complexity, Fisher-Shannon information plane, Shannon entropy power, Fisher information measure, statistical complexity, non-linear time series, dynamical behavior characterization, high frequency wind speed

## 1. INTRODUCTION

The ubiquity and extensive growth of available temporal data requires the development of reliable techniques to extract knowledge from them and to understand multifaceted time-dependent phenomena. Over the last decades, an increasing attention was payed toward the use of Fisher-Shannon information as a measure to characterize the complexity and non-stationarity of non-linear time series. Originally proposed for statistical estimation purposes (Fisher, 1925), the Fisher information measure (FIM) has been extensively used in theoretical physics (Frieden, 1990). FIM and Shannon entropy power (SEP) (Shannon, 1948) are closely related, as shown by information theory (Dembo et al., 1991; Cover and Thomas, 2006). The Fisher-Shannon complexity (FSC)—the FIM and SEP product—was proposed as a possible definition of atom complexity (Angulo et al., 2008; Esquivel et al., 2010).

Following Frieden work, FIM has found applications in non-linear time-series analysis. Martin et al. (1999) analyzed complex non-stationary electroencephalographic signals and showed that FIM can have better discrimination performance than Shannon entropy. FIM was also used to detect behavior changes of dynamical systems (Martin et al., 2001). Vignat and Bercher (2003) showed that a joint analysis of both SEP and FIM can be required to perform effective discrimination of non-stationary signals.

The Fisher-Shannon method has been used to analyse complex dynamical processes in geophysics. Discrimination between the electric and magnetic components of magnetotelluric signals is performed in Telesca et al. (2011). Tsunamigenic and non-tsunamigenic earthquakes were efficiently separated in the Fisher-Shannon information plane, using FSC (Telesca et al., 2013). Micro-tremors time series were identified depending on the soil characteristics of the measurement sites (Telesca et al., 2015b). Telesca et al. (2015a) proposed a classifier of (non-)tsunamigenic potential of earthquake build on several time series features, including FIM, SEP, FSC. Finally, FIM was also used dynamically with sliding window techniques in order to study precursory patterns in seismology (Telesca et al., 2009b) and volcanology (Telesca et al., 2010).

Many environmental processes have also been studied using the Fisher-Shannon method. Lovallo et al. (2013) and Pierini et al. (2011) studied climatic regimes identification in rainfall time series. Hydrological regimes discrimination have also been investigated (Pierini et al., 2015). Analyzing remotely sensed sea surface temperature, Pierini et al. (2016) have shown that the Fisher-Shannon method is able to clearly identify the Brazil-Malvinas Confluence Zone, which is known to be one of the most energetic area of oceans. Telesca and Lovallo (2011) analyzed more than 10 years of hourly wind speed data in the Fisher-Shannon information plane. The same authors studied yearly variation of the FIM, the SEP and the FSC on wind measurements (Telesca and Lovallo, 2013). Guignard et al. (2019b) have found correlations between daily variance of temperature and daily FSC of high-frequency wind speed records in urban area. Authors have also pointed out relationships between Fisher-Shannon analysis of wind speed daily means and topographical features—height and slope—in complex mountainous regions (Guignard et al., 2019a). Telesca et al. (2009a) discriminated some pollutants, including cadmium, iron, and lead, in the Fisher-Shannon plane. Similarly, Amato et al. (2020) have shown a relationship between the Fisher-Shannon analysis outputs of three air pollutants—Nitrogen dioxide, Ground level ozone and Particulate Matter—and measurement location in term of land use and of anthropogenic sources of pollutant emission.

The research involving Fisher-Shannon method is rather scattered and comes from various fields, e.g., information theory, physics, dynamical systems, machine learning, and statistics. Therefore, the present paper contributes to the methodological studies on Fisher-Shannon information measures along with some applications.

The main objectives of this research can be summarized as follows:

- discussing the state-of-the-art of Fisher-Shannon information measures and their applications,
- identifying FSC as a sensitivity measure of the SEP and as a scale-independent non-Gaussianity measure of data,
- presenting some new theoretical results on FIM and SEP,
- developing operational FIM and SEP tools for the nonlinear time-series analysis,
- demonstrating through two case studies, based on simulated (chaotic) and real data (high frequency wind speed measurements), the efficiency and usefulness of the proposed methods.

The remainder of the paper is organized as follows. Concepts of Fisher-Shannon analysis, including SEP, FIM, FSC, and information plane, are presented and reviewed in section 2. Section 3 provides analytical formula for such quantities in the particular cases of random variables following some positive skewed distributions, namely Gamma, Weibull, and log-normal ones. Then, a non-parametric kernel based estimation of the density function—for which Python and R packages are proposed—is presented in section 4. Experiments on simulated and real-world data are performed in section 5. Finally, section 6 concludes the paper.

## 2. FISHER-SHANNON ANALYSIS

### 2.1. Shannon Entropy Power and Fisher Information Measure

Let us consider a univariate continuous random variable $X$ with its probability density function (PDF) $f(x)$, which is supposed to be sufficiently regular for the exposition of our purpose. Its *differential entropy* (Cover and Thomas, 2006) is defined as

$$H_X = \mathbb{E}\left[-\log f(X)\right] = -\int f(x) \log f(x)\, dx. \quad (1)$$

For example, if $X$ is a centered Gaussian random variable of variance $\sigma^2$, a direct computation gives $H_X = \frac{1}{2}\log(2\pi e \sigma^2)$. However, it will be more convenient to work with the following quantity, called the *Shannon Entropy Power* (SEP) (Dembo et al., 1991),

$$N_X = \frac{1}{2\pi e} e^{2H_X}, \quad (2)$$

which is a strictly monotonically increasing transformation of $H_X$. The SEP is constructed such that in the Gaussian case we have $N_X = \sigma^2$. Very often, entropies $H_X$ and $N_X$ are interpreted as global measures of disorder / uncertainty / spread of $f(x)$. The higher the entropy, the higher the disorder.

The *Fisher Information Measure* (FIM) (Vignat and Bercher, 2003), also known as the *Fisher information of X with respect to a scalar translation parameter* (Dembo et al., 1991), is defined as

$$I_X = \mathbb{E}\left[\left(\frac{\partial}{\partial x}\log f(X)\right)^2\right] = \int \frac{\left[\frac{\partial}{\partial x}f(x)\right]^2}{f(x)}\, dx. \quad (3)$$

This quantity should not be confused with the Fisher information of a distribution parameter. In particular, the derivative of the log-density is relative to $x$ and not to some parameter. However, the FIM is equivalent to the Fisher information of a location parameter of a parametric distribution (Cover and Thomas, 2006). Under mild regularity conditions, one has the following alternative formulation (Lehmann, 1999),

$$I_X = \mathbb{E}\left[-\frac{\partial^2}{\partial x^2}\log f(X)\right]. \quad (4)$$

The quantity $I_X$ is sometimes interpreted as a measure of order / organization / narrowness of $X$. If $X$ is Gaussian, $I_X = 1/\sigma^2$. It should be noted that $H_X, N_X$, and $I_X$ only depend on the distribution $f(x)$.

## 2.2. Properties

The SEP and the FIM respect several properties. First, both quantities are positive. It is also easy to see the *scaling properties* of the SEP and the FIM (Rioul, 2011),

$$N_{aX} = a^2 N_X,$$
$$I_{aX} = a^{-2} I_X. \quad (5)$$

for any real number $a \neq 0$, by change of variable. Notice also that the SEP and the FIM are invariant under additive deterministic constant, by the same argument. Harder to show are the *entropy power inequality* (Dembo et al., 1991) and its dual the *Fisher information inequality* (Zamir, 1998),

$$N_{X+Y} \geq N_X + N_Y, \quad (6)$$
$$I_{X+Y}^{-1} \geq I_X^{-1} + I_Y^{-1}, \quad (7)$$

for a random variable $Y$ independent of $X$, with equality if $X$ and $Y$ are Gaussian.

Moreover, several relationships show that the FIM closely interact with the SEP and the differential entropy. Let $Z$ be a random variable independent of $X$ with finite variance $\sigma_Z^2$. The *de Bruijn's identity* (Cover and Thomas, 2006; Rioul, 2011) states that

$$\frac{d}{dt} H_{X+\sqrt{t}Z}\bigg|_{t=0} = \frac{1}{2} \sigma_Z^2 I_X, \quad (8)$$

i.e., the variation of the differential entropy of a perturbed $X$ is proportional to $I_X$. Therefore, a possible interpretation of the FIM is that it quantifies the sensitivity of $H_X$ to a small independent additive perturbation $Z$. Using the entropy power inequality (6) and de Bruijn identity (8), one can show the *isoperimetric inequality for entropies*,

$$N_X I_X \geq 1, \quad (9)$$

with equality if and only if $X$ is Gaussian. The proof and the nomenclature motivation of equation (9) can be found in Dembo et al. (1991), where a remarkable analogy is done with geometry. This shows that SEP and FIM are intimately interlinked.

## 2.3. Fisher-Shannon Complexity

The joint FIM/SEP analysis has been used as a statistical complexity measure, albeit there is no clear consensus about the definition of signal complexity (Esquivel et al., 2010). The *Fisher-Shannon Complexity* (FSC) is defined as $C_X = N_X I_X$ (Angulo et al., 2008). From the scaling properties (5), it is easy to show that the FSC is constant under scalar multiplication and addition. In particular, normalization or standardization of $X$ has no effect on the FSC. Additionally, the isoperimetric inequality for entropies (9) states that $C_X \geq 1$, with equality if and only if $X$ is Gaussian. An interpretation of this quantity is the following.

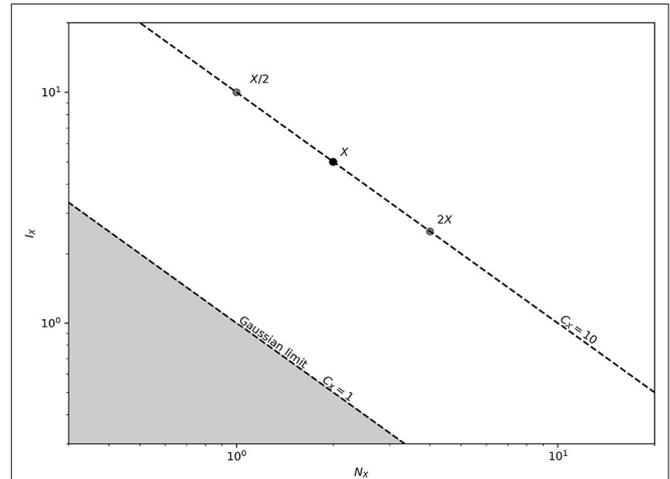

**FIGURE 1** | The Fisher-Shannon information plane with a random variable $X$ of FSC equal to 10. Scalar multiplication of $X$ corresponds to a displacement along the iso-complex curve passing through $X$. The unreachable points are in gray. Note the logarithmic scale.

If $Z$ is independent of $X$ and has a finite variance $\sigma_Z^2$, one obtains the following relationship by using the de Bruijn identity (8),

$$\frac{d}{dt} N_{X+\sqrt{t}Z}\bigg|_{t=0} = 2N_X \frac{d}{dt} H_{X+\sqrt{t}Z}\bigg|_{t=0} = \sigma_Z^2 N_X I_X = \sigma_Z^2 C_X.$$

Hence, the FSC can be interpreted as a sensitivity measure of $N_X$ to a small independent additive perturbation.

## 2.4. Fisher-Shannon Information Plane

The PDF of $X$ can be analyzed displaying the SEP and FIM within the so-called Fisher-Shannon Information Plane (FSIP), see **Figure 1** (Vignat and Bercher, 2003). Although standard linear scale plot are very often used for the FSIP in the literature, a log-log plot is more adequate in practice. In the FSIP, the only reachable values are in the set $\mathcal{D} = \{(N_X, I_X) \in \mathbb{R}^2 | N_X > 0, I > 0 \text{ and } N_X I_X \geq 1\}$, due to (9). Vignat and Bercher (2003) showed that for any point $(N, I) \in \mathcal{D}$, it exists a random variable $X$ (from an exponential power distribution) such that $N_X = N$ and $I_X = I$.

A curve in $\mathcal{D}$ is said to be *iso-complex* if the FSC along the curve is constant. As $C_X$ is constant up to a multiplicative factor $a \neq 0$, and looking up at the scaling properties (5), one can move on any iso-complex curve by varying $a$. **Figure 1** shows the iso-complex curve of complexity $C_X = 10$ as an example. The boundary of $\mathcal{D}$ is the iso-complex curve with FSC equal to 1, and is reached if and only if $X$ is Gaussian, as states by (9). On this boundary, the standard deviation $\sigma$ (which plays the role of the scaling parameter in the Gaussian case) is equivalent to the multiplicative factor $a$. Hence, while a point in the FSIP is described by $(N_X, I_X)$, one can also describe it by $(a, C_X)$. In the light of this, one can also think of FSC as a scale-independent measure of non-Gaussianity of $X$.

## 3. ANALYTICAL SOLUTIONS FOR SOME DISTRIBUTIONS

In this section, we propose analytical formulas for the SEP, FIM and FSC for several parametric distributions. They can be used directly for parametric estimations. Vignat and Bercher (2003) obtained analogous results for the Student's t-distribution and the exponential power distribution (also known as generalized Gaussian distribution). The Gaussian case was already presented in section 2 as an example.

The differential entropy of the distributions proposed in this section have been computed by Lazo and Rathie (1978), from which the SEP is directly obtained. However, to our knowledge, the FIM-based calculations for Gamma, Weibull and log-normal distributions were never presented. Proofs can be found in the **Appendix**.

### 3.1. Gamma Distribution

The PDF of a Gamma random variable $X$ is given by

$$f(x) = f(x; \theta, k) = \frac{x^{k-1} e^{-\frac{x}{\theta}}}{\theta^k \Gamma(k)}, \qquad \text{for } x \geq 0,$$

and $f(x) = 0$, for $x < 0$, where $\Gamma$ denotes the gamma function and $\theta, k > 0$ are, respectively, the scale and shape parameters.

Proposition 1. *The SEP of the Gamma distribution with scale $\theta > 0$ and shape $k > 0$ is*

$$N_X(\theta, k) = \frac{\theta^2 \Gamma^2(k)}{2\pi e} e^{2[(1-k)\psi(k)+k]},$$

*where $\psi$ is the digamma function.*

*The FIM and the FSC of the Gamma distribution with scale $\theta > 0$ and shape $k > 2$ are, respectively,*

$$I_X(\theta, k) = \frac{1}{(k-2)\theta^2},$$

$$C_X(k) = \frac{\Gamma^2(k)}{2\pi e(k-2)} e^{2[(1-k)\psi(k)+k]}.$$

### 3.2. Weibull Distribution

The PDF of a Weibull random variable is

$$f(x) = f(x; \mu, \lambda, k) = \frac{k}{\lambda}\left(\frac{x-\mu}{\lambda}\right)^{k-1} e^{-\left(\frac{x-\mu}{\lambda}\right)^k}, \qquad \text{for } x \geq 0,$$

and $f(x) = 0$, for $x < 0$, where $\mu$ is the location parameter, $\lambda > 0$ is the scale parameter and $k > 0$ is the shape parameter.

Proposition 2. *The SEP of the Weibull distribution with location $\mu$, scale $\lambda > 0$ and shape $k > 0$ is*

$$N_X(\lambda, k) = \frac{(1-\alpha)^2 \lambda^2 e}{2\pi} e^{2\alpha\gamma},$$

*where $\alpha = \frac{k-1}{k}$ and $\gamma$ is the Euler-Mascheroni constant.*

*The FIM and the FSC of the Weibull distribution with location $\mu$, scale $\lambda > 0$ and shape $k > 2$ are, respectively*

$$I_X(\lambda, k) = \frac{\alpha^2}{(1-\alpha)^2 \lambda^2} \Gamma(2\alpha - 1),$$

$$C_X(k) = \frac{\alpha^2 e}{2\pi} \Gamma(2\alpha - 1) e^{2\alpha\gamma}.$$

### 3.3. Log-Normal Distribution

The log-normal PDF with parameters $\mu$ and $\sigma > 0$ is

$$f(x) = f(x; \mu, \sigma) = \frac{1}{x\sigma\sqrt{2\pi}} e^{-\frac{(\log x - \mu)^2}{2\sigma^2}}, \qquad \text{for } x > 0,$$

and $f(x) = 0$, for $x \leq 0$.

The notation of the parameters $\mu$ and $\sigma$ are motivated by the fact that the logarithm of a log-normal random variable follows a normal distribution of mean $\mu$ and variance $\sigma^2$. However, $\mu$ and $\sigma$ play, respectively, the role of the scale parameter and the shape parameter for the log-normal distribution.

Proposition 3. *The SEP, the FIM and the FSC of the log-normal distribution with $\mu$ and $\sigma > 0$ are given by*

$$N_X(\mu, \sigma) = \sigma^2 e^{2\mu},$$

$$I_X(\mu, \sigma) = \left(1 + \frac{1}{\sigma^2}\right) e^{2(\sigma^2 - \mu)},$$

$$C_X(\sigma) = (1 + \sigma^2) e^{2\sigma^2}.$$

## 4. DATA DRIVEN NON-PARAMETRIC ESTIMATION

Complex real-world data sets rarely follow parametric distributions. Providing enough data, it is also possible to carry out Fisher-Shannon analysis with a non-parametric estimation of density, which release parametric assumptions on the distribution (Telesca and Lovallo, 2017). In this paper, *integral estimates* of the SEP and the FIM are considered, which consist of substituting the kernel density estimators (KDE) of both $f(x)$ and its derivative in the integral forms of (1) and (3) (Bhattacharya, 1967; Dmitriev and Tarasenko, 1973; Prakasa Rao, 1983; Györfi and van der Meulen, 1987; Joe, 1989). Python and R implementations of this section content are proposed, see the software availability section at the end of this paper.

Following (Wand and Jones, 1994), let $X_1, \ldots, X_n$ be a random sample of size $n$ from a PDF $f(x)$. Consider also the *kernel* $K(u)$, a bounded PDF which is symmetric around zero, has a finite fourth moment and is differentiable. The KDE of $f(x)$ is

$$\hat{f}_h(x) = \frac{1}{nh} \sum_{i=1}^{n} K\left(\frac{x - X_i}{h}\right), \qquad (10)$$

where $h > 0$ is the *bandwidth* parameter. In this paper, the Gaussian kernel defined by $K(u) = (2\pi)^{-1/2} \exp(-u^2/2)$ is used and the estimator (10) becomes

$$\hat{f}_h(x) = \frac{1}{\sqrt{2\pi} nh} \sum_{i=1}^{n} \exp\left\{-\frac{1}{2}\left(\frac{x - X_i}{h}\right)^2\right\}.$$

The integral estimate of (2) is

$$\hat{N}_X = \frac{1}{2\pi e} \exp\left\{-2 \int \hat{f}_h(x) \log \hat{f}_h(x)\, dx\right\}.$$

Let us note $f'$, the derivative of $f$ with respect to $x$. Usually, $f'$ is estimated by $\hat{f}'_h$. With the Gaussian kernel we obtain

$$\hat{f}'_h(x) = \frac{-1}{\sqrt{2\pi}\, nh^3} \sum_{i=1}^{n} (x - X_i) \exp\left\{-\frac{1}{2}\left(\frac{x - X_i}{h}\right)^2\right\}.$$

Then, the integral estimate of (3) is

$$\hat{I}_X = \int \frac{\left(\hat{f}'_h(x)\right)^2}{\hat{f}_h(x)}\, dx.$$

The FSC is estimated by multiplying $\hat{N}_X$ by $\hat{I}_X$.

Several techniques exist in order to automatize the bandwidth choice (Wand and Jones, 1994). In the following, the 2-stages direct plug-in method (Sheather and Jones, 1991) is used. This method estimates the optimal bandwidth regarding the asymptotic mean integrated squared error of $\hat{f}_h$. The interested reader can found further technical details in (Wand and Jones, 1994) and (Sheather and Jones, 1991).

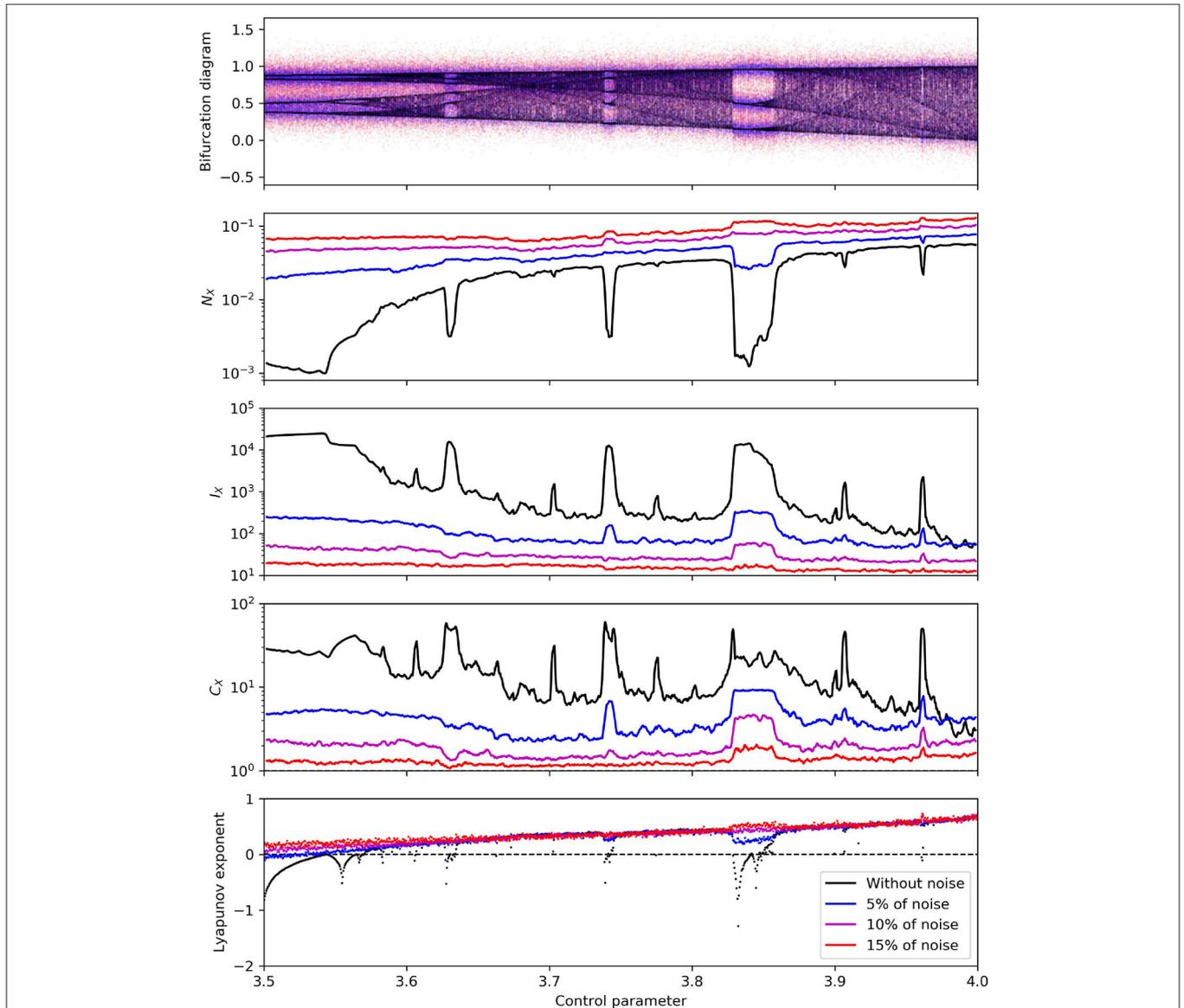

**FIGURE 2** | Logistic map with different level of noise. From top to bottom : bifurcation diagram, SEP, FIM, FSC, and Lyapunov exponent sliding windows. Note the logarithmic scale on the y-axis for SEP, FIM, and FSC.

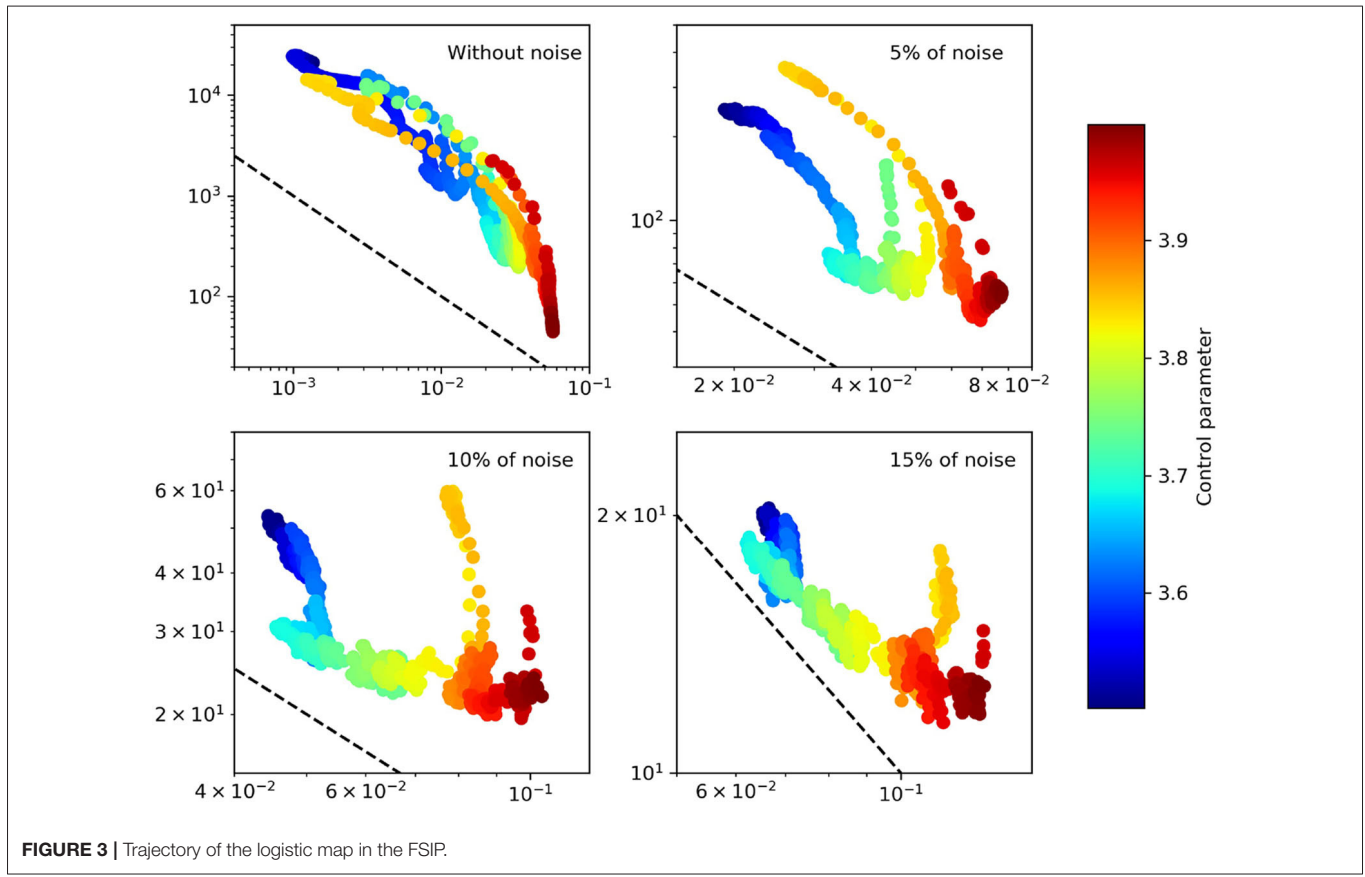

**FIGURE 3** | Trajectory of the logistic map in the FSIP.

## 5. CASE STUDIES

In this section we explore two applications of SEP, FIM and FSC to time series. First, a synthetic experiment is used to show the usefulness of the method in detecting the dynamical behavior of chaotic systems. Then, an example of application of the proposed method to real complex environmental data is discussed.

### 5.1. Logistic Map

A synthetic experiment is designed to investigate how SEP, FIM, and FSC can be used to detect behavioral changes in non-linear dynamical systems. In the present research, the well-known logistic map is considered as a benchmark simulated case study, that illustrates and helps to understand important features of the considered measures.

Following the experiment proposed by Martin et al. (2001), the *logistic map* defined by

$$x_{n+1} = cx_n(1-x_n), \qquad x_0 \in [0,1], \quad c \in [0,4],$$

where $c$ is the control parameter, is analyzed using sliding window technique. Analysis within the sliding window pursues the goal of revealing dynamical evolution of properties of time series like Gaussianity and non-stationarity.

The sequence $(x_n)$ is computed up to $n = 1,000$ for $c \in [3.5, 4]$. Centered Gaussian noise with different level of variance, 0.05, 0.10, 0.15, is added to $x_n$. The well-known bifurcation diagram of the logistic map is displayed in **Figure 2**. The SEP, FIM, and FSC are computed on data included in the overlapping windows of width $2.5 \cdot 10^{-3}$ along the control parameter, and the results are shown in the same figure. The Lyapunov exponent is also added for comparison reasons (Kantz and Schreiber, 2004). The results are also displayed in the FSIP, see **Figure 3**.

Analyzing the results obtained from the data without noise, it is easy to see how the SEP, FIM and FSC peak occurrences correspond to dynamic changes shown by the bifurcation diagram and the Lyapunov exponent. With the logarithmic scale on the y-axis, the behavior of the SEP is somewhat symmetric to the behavior of the FIM, i.e., the FIM seems to be inversely proportional to the SEP. However, this is not exactly the case, otherwise the FSC would be constant. In some sense, the perturbations in the FSC reflect the departure from the inverse proportionality between the SEP and the FIM. In the FSIP, perfect inverse proportionality corresponds to iso-complex curves. Indeed, the trajectory of the logistic map in the FSIP is stretched along iso-complex curves, see **Figure 3**.

Adding noise shows that most of the peaks become undetectable, see **Figure 2**. However, FSC seems to be the measure which suffers the least to noise in data. Note also, that FIM is less impacted than SEP. The noise effect is more interesting in the FSIP, see **Figure 3**. While the uncorrupted data is quite hard to interpret due to the superposition of the trajectory with itself, adding some noise seems to clarify complexity and trajectory behaviors in the FSIP. Noise stimulates the emergence of protuberances roughly corresponding to "islands of stability" of the (uncorrupted) bifurcation diagram, where Lyapunov exponent is negative. This emergence is due to the fact that FIM is less impacted than SEP, as it was seen above.

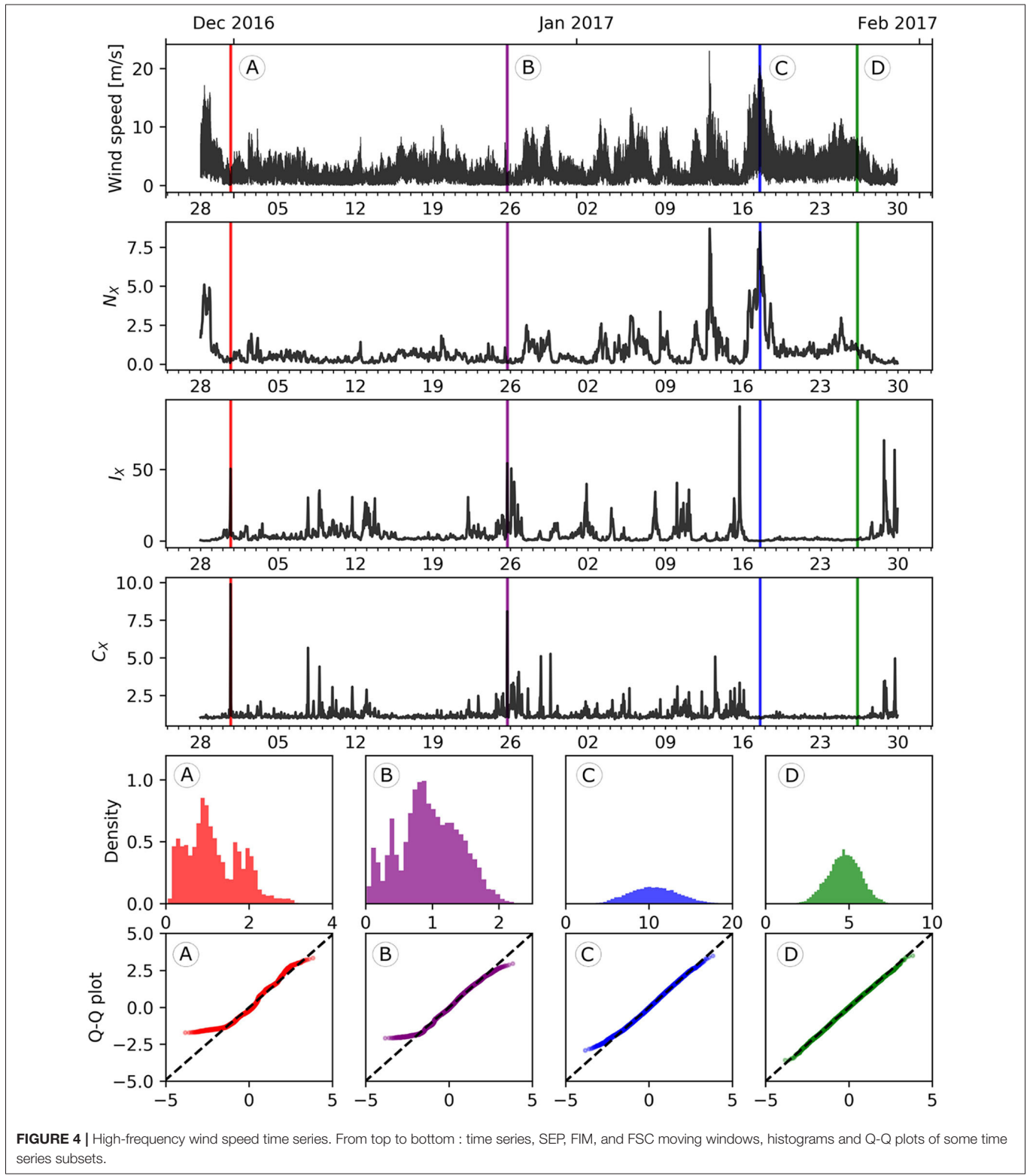

**FIGURE 4** | High-frequency wind speed time series. From top to bottom : time series, SEP, FIM, and FSC moving windows, histograms and Q-Q plots of some time series subsets.

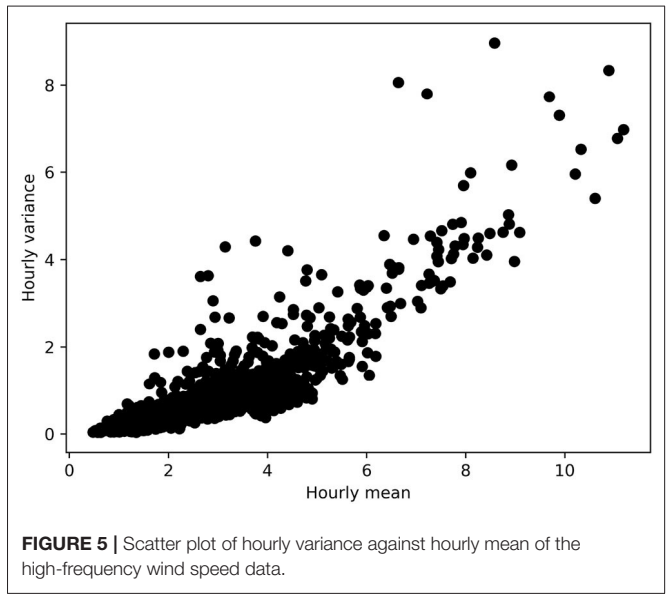

FIGURE 5 | Scatter plot of hourly variance against hourly mean of the high-frequency wind speed data.

## 5.2. Application to High Frequency Wind Data

The Fisher-Shannon information method can find a wide application in the geo-environmental domains. In the present section, we demonstrate how they can be applied to retrieve relevant knowledge from environmental time series. Specifically, high frequency wind speed data are analyzed. The time series consists of 1Hz frequency wind speed data, from 28 November 2016 to 29 January 2017 (**Figure 4**). The data (motus.epfl.ch) were measured at 25.5 m above the ground by a sensor which is placed on meteorological mast located on the campus of the Ecole Polytechnique Fédérale de Lausanne (EPFL), Switzerland. Notice that the mast is surrounded by a building layout of 10 m average height. More information on these measurements can be found in Mauree et al. (2017a,b).

The Fisher-Shannon quantities are computed with non-overlapping moving windows of 1 hour width along the time axis. Globally, all quantities vary with time, indicating non-stationarity, see **Figure 4**. The SEP seems to roughly replicate the behavior of the original time series. This is due to a proportional effect between the mean and the variance of the data, as shown in **Figure 5**. As for the logistic map case, the FIM is roughly inversely proportional to the SEP (not shown in logarithmic scale). The FSC is close to 1 during long period of time, e.g., between the 17th and the 27th January 2017. This should indicate a local behavior of wind speed close to a Gaussian one. During these periods, wind speed is not necessarily calm, e.g., the 17th January. Conversely, The FSC also exhibits some peaks where wind speed is rather low, which should indicate a more complex distribution of the data.

To verify this, a closer exploration of the data is required. To this aim, we considered four subsets of 3 hours length, denoted by A, B, C, and D and represented on **Figure 4** by color red, purple, blue, and green, respectively. Histograms and quantile-quantile (Q-Q) plots of these data subsets are also plotted with the corresponding colors. The subset D is chosen during the period of almost unitary FSC. The corresponding histogram and Q-Q plot confirms the very-close-to-Gaussian behavior of the data. The subset C is also chosen with a FSC close to 1, but centered on the maximum of SEP of the 17th January 2017 which corresponds also to a high wind speed activity. The histogram shows again a distribution close to a Gaussian one, but with a higher variance than C. This was an expected output, since for Gaussian distribution the SEP equals to the variance and C was chosen with a high SEP. The Q-Q plot shows little departure for the left tail, but the data are still relatively close to what was expected. The subset B is centered on a peak of FSC. The histogram shows a distribution which is very far from Gaussianity. It is clearly asymmetric and has at least two modes—maybe three. The Q-Q plot shows a strong departure from the Gaussian distribution, especially on the left tail. The subset A is centered on the highest FSC value. Its histogram shows three—maybe four—modes. The corresponding Q-Q plot shows how for this subset data are even farther from Gaussianity than for the previous subset.

These results show the high complexity of these data, whose behavior can rapidly change locally in time or even during calm weather. Further analysis on a larger set of these measurements using the FSC can be found in Guignard et al. (2019b), where authors analyzed wind speed and temperature data gathered by sensors similar to the one used to collect the data analyzed in this section, fixed along a mast located in a urban canyon. A FSC analysis suggested different wind dynamics induced by the building layout. The daily variation of temperature was also found to be an important predictor for high-frequency wind speed daily complexity. Moreover, FSC was used to show that wind speed and height are related by a non-linear relationship. More generally, this demonstrates the high versatility of analysis based on Fisher-Shannon information, which had yield numerous and various insight on these data.

## 6. CONCLUSIONS

The paper discusses the Fisher-Shannon information method as an effective data exploration tool able to give diverse insights into complex non-stationary time series. The Fisher-Shannon method was presented in a unified framework and new interpretations of FSC were pointed out. In particular, the FSC was identified as a sensitivity measure of the SEP and as a scale-independent non-Gaussianity measure, which both provide interpretation of this quantity as well. The detection of potential Gaussian behavior in the data was successfully showed on high-frequency wind speed data.

In the methodological part of the paper, FIM and FSC were computed in closed forms for several parametric distributions which are widely used in geo-environmental data analyses. Theoretical formulas for other random variables can be derived depending on the problem at hand. Furthermore, it was also shown—by injecting noise in the logistic map—how these information measures can be used to detect potential dynamic changes in a quite robust manner—especially with FSC. While SEP, FIM and FSC were presented as versatile information-based exploratory tools, they can also be used as time series discrimination or, more generally, to generate time series features for clustering, modeling, and forecasting.

The Fisher-Shannon method has been widely used in geosciences, as shown in the first part of this paper. However, according to our opinion, its full potential is still unexploited and underestimated. To simplify the access to this method for environmental data analysis and foster reproducibility, open source libraries written in R and Python for the computation of the three measures via a non-parametric kernel density estimation are provided.

From a theoretical point of view, future studies should involve generalization of the Fisher-Shannon method to the multivariate case. Several numerical investigations could be carried out for the KDE of the FIM. In particular, other estimates could be provided by re-substitution techniques as with entropy. Optimal bandwidth choice regarding to asymptotic mean squared error of FIM—or even FSC—could be derived. More practically, a challenging exploratory analysis of spatio-temporal data is planned.

## SOFTWARE AVAILABILITY

A Python package is proposed on PyPI and GitHub (https://github.com/fishinfo/FiShPy), as well as an R package available on CRAN and GitHub (https://github.com/fishinfo/FiSh). They allow interested users to compute non-parametric estimation of the SEP, FIM and FSC.

## DATA AVAILABILITY STATEMENT

The data presented in this article are not readily available as the authors are not their owners. Requests to access the data should be directed to the responsible parties of the MoTus experiment (motus.epfl.ch), Dasaraden Mauree (dasaraden.mauree@epfl.ch).

## AUTHOR CONTRIBUTIONS

FG conceived the main conceptual ideas, conduct investigations, developed the theoretical formalism, performed the calculations, interpreted the computational results, and wrote the original draft. FG and ML developed Python and R packages. FG, FA, and MK wrote the final version of the paper. MK carried out the supervision, project administration, and funding acquisition. All authors discussed the results, provided critical feedback, commented, reviewed and edited the original manuscript, and gave final approval for publication.

## FUNDING

FG and MK acknowledge the support of the National Research Programme 75 Big Data (PNR75) of the Swiss National Science Foundation (SNSF), project no. 167285.

## ACKNOWLEDGMENTS

The authors are grateful to Dasaraden Mauree from the Ecole Polytechnique Fédérale de Lausanne, Switzerland, for providing the high-frequency wind speed data, and to Sylvain Robert for analytical computation checking. They also thank Luciano Telesca from Institute of Methodologies for Environmental Analysis of the Italian National Research Council for profitable discussions, and the reviewers for their constructive comments, which contributed to improve the paper.

# 7. APPENDIX

The differential entropy $H_X$ for Gamma, Weibull and log-normal distributions can be found in (Lazo and Rathie, 1978) and (Cover and Thomas, 2006). The SEP is simply a non-linear transform of $H_X$.

*Proof of proposition 1:* Computing the second derivative of $\log f(x)$, one has

$$\frac{\partial^2}{\partial x^2} \log f(x) = -\frac{k-1}{x^2},$$

and then, using (4), the variable change $x = \theta y$ and the properties of the Gamma function,

$$\begin{aligned}
I_X &= (k-1)\mathbb{E}[X^{-2}] \\
&= \frac{k-1}{\theta^k \Gamma(k)} \int_0^\infty x^{k-3} e^{-\frac{x}{\theta}} dx \\
&= \frac{k-1}{\theta^2 \Gamma(k)} \int_0^\infty y^{k-3} e^{-y} dy \\
&= \frac{(k-1)\Gamma(k-2)}{\theta^2 (k-1)(k-2)\Gamma(k-2)},
\end{aligned}$$

yielding the FIM for the Gamma distribution. The FSC is directly obtained by multiplying the SEP and the FIM. □

*Proof of proposition 2:* Starting from the Weibull PDF, one has

$$\frac{\partial^2}{\partial x^2} \log f(x) = -\frac{k-1}{(x-\mu)^2} - \frac{k(k-1)}{\lambda^k}(x-\mu)^{k-2},$$

and with the variable change $y = (\frac{x-\mu}{\lambda})^k$,

$$\begin{aligned}
I_X &= (k-1)\mathbb{E}\left[(X-\mu)^{-2}\right] + \frac{k(k-1)}{\lambda^k}\mathbb{E}\left[(X-\mu)^{k-2}\right] \\
&= \frac{k(k-1)}{\lambda^3} \left[ \int_0^\infty \left(\frac{x-\mu}{\lambda}\right)^{k-3} e^{-(\frac{x-\mu}{\lambda})^k} dx \right. \\
&\quad + k \left. \int_0^\infty \left(\frac{x-\mu}{\lambda}\right)^{2k-3} e^{-(\frac{x-\mu}{\lambda})^k} dx \right]
\end{aligned}$$

$$\begin{aligned}
&= \frac{k-1}{\lambda^2} \left[ \int_0^\infty y^{-\frac{2}{k}} e^{-y} dy + k \int_0^\infty y^{1-\frac{2}{k}} e^{-y} dy \right] \\
&= \frac{k-1}{\lambda^2} \left[ \Gamma\left(1 - \frac{2}{k}\right) + k\Gamma\left(2 - \frac{2}{k}\right) \right] \\
&= \frac{k-1}{\lambda^2} \left[ 1 + k\left(1 - \frac{2}{k}\right) \right] \Gamma\left(1 - \frac{2}{k}\right) \\
&= \frac{(k-1)^2}{\lambda^2} \Gamma\left(1 - \frac{2}{k}\right).
\end{aligned}$$

□

*Proof of proposition 3:* The second derivative of $\log f(x)$ is

$$\frac{\partial^2}{\partial x^2} \log f(x) = \frac{\log x - \mu + \sigma^2 - 1}{\sigma^2 x^2},$$

and using the variable change $y = \log x - \mu$, one have

$$\begin{aligned}
I_X &= \frac{1}{\sigma\sqrt{2\pi}} \int_0^\infty \frac{1 - \sigma^2 - (\log x - \mu)}{\sigma^2 x^3} e^{-\frac{(\log x - \mu)^2}{2\sigma^2}} dx \\
&= \frac{1}{\sigma\sqrt{2\pi}} \int_{-\infty}^\infty \frac{1 - \sigma^2 - y}{\sigma^2} e^{-\frac{y^2}{2\sigma^2} - 2y - 2\mu} dy.
\end{aligned}$$

Note that

$$-\frac{y^2}{2\sigma^2} - 2y - 2\mu = -\frac{(y + 2\sigma^2)^2}{2\sigma^2} + 2(\sigma^2 - \mu).$$

Using this and the definition of a Gaussian distribution $\mathcal{N}(-2\sigma^2, \sigma)$,

$$\begin{aligned}
I_X &= \frac{1}{\sigma\sqrt{2\pi}} \int_{-\infty}^\infty \left(\frac{1-\sigma^2}{\sigma^2} - \frac{y}{\sigma^2}\right) e^{-\frac{(y+2\sigma^2)^2}{2\sigma^2} + 2(\sigma^2 - \mu)} dy \\
&= \frac{e^{2(\sigma^2 - \mu)}}{\sigma^2} \left[ \frac{1-\sigma^2}{\sigma\sqrt{2\pi}} \int_{-\infty}^\infty e^{-\frac{(y+2\sigma^2)^2}{2\sigma^2}} dy \right. \\
&\quad \left. - \frac{1}{\sigma\sqrt{2\pi}} \int_{-\infty}^\infty y e^{-\frac{(y+2\sigma^2)^2}{2\sigma^2}} dy \right] \\
&= \frac{e^{2(\sigma^2 - \mu)}}{\sigma^2} \left[1 - \sigma^2 + 2\sigma^2\right] \\
&= \frac{1 + \sigma^2}{\sigma^2} e^{2(\sigma^2 - \mu)},
\end{aligned}$$

and the FIM is obtained. The FSC is

$$C_X = (1 + \sigma^2) e^{2\mu + 2(\sigma^2 - \mu)} = (1 + \sigma^2) e^{2\sigma^2}.$$

□